\documentclass{ws-procs9x6}

\newcommand{\lsim}{\, \, \raisebox{-0.8ex}{$\stackrel{\textstyle <}{\sim}$ }}

\begin{document}

\title{Working Group Summary: Chiral Dynamics in 
Few-Nucleon Systems}

\author{H.-W.~HAMMER}

\address{HISKP (Theorie), Universit\"at Bonn,
Nu\ss allee 14-16, D-53115 Bonn, Germany\\
E-mail: hammer@itkp.uni-bonn.de}

\author{N.~KALANTAR-NAYESTANAKI}

\address{Kernfysisch Versneller Instituut (KVI), University of Groningen,\\
Groningen, The Netherlands\\
E-mail: nasser@kvi.nl}

\author{D.~R.~PHILLIPS}

\address{Department of Physics and Astronomy, Ohio University,
Athens, OH 45701, USA\\
E-mail: phillips@phy.ohiou.edu}

\begin{abstract}
We summarize the findings of our Working Group, which discussed
progress in the understanding of Chiral Dynamics in the $A=2$, $3$,
and $4$ systems over the last three years.  We also identify
key unresolved theoretical and experimental questions in
this field.
\end{abstract}

\keywords{Chiral Dynamics; Few-Nucleon Systems}

\bodymatter

\section{Introduction}\label{sec-intro}

This year is the tenth anniversary of the first quantitative treatment
of the nuclear force based on chiral perturbation theory ($\chi$PT):
the seminal paper of Ord\'o\~nez {\it et al.}~\cite{Or96}. This paper
employed Weinberg's proposal to expand the $NN$ potential up to a
given chiral order:
\begin{equation}
V=V^{(0)} + V^{(2)} + V^{(3)} + V^{(4)} + \ldots,
\label{eq:V}
\end{equation}
and then use the resulting potential in the Schr\"odinger equation
and obtain the $NN$ wave function~\cite{We91}. However, in
that same year Kaplan, Savage, and Wise raised serious questions about
the consistency of such an approach~\cite{Ka96}.

The paper of Ord\'o\~nez {\it et al.} analyzed $NN$ scattering data
using an $NN$ potential which included all mechanisms up to
next-to-next-to-leading order (N$^2$LO). This has now been extended
to one higher order~\cite{EM03,EMG05}, and the result is an
energy-independent potential that describes $NN$ data with an accuracy
comparable to that of the ``high-quality'' $NN$ potentials.
However,
questions remain about the consistency of the power counting used to
derive this N$^3$LO potential~\cite{ES02,Nogga:2005hy,Birse:2005um}.
These questions were discussed in a special ``panel'' format within
our Working Group. The conclusions of this panel are summarized in
Section~\ref{sec-NN}.

At the same time lattice simulations have made significant strides in
connecting these chiral potentials to QCD itself. The work of the
NPLQCD collaboration is also discussed in Sec.~\ref{sec-NN}, and it
shows that $NN$ scattering lengths are natural at quark masses
corresponding to $m_\pi \geq 350$ MeV~\cite{Beane:2006mx}. Chiral
extrapolation from the physical point to these higher values of
$m_\pi$ shows that the numbers obtained in the simulations are
consistent with the experimentally measured ``unnaturally large''
($|a|\gg 1/m_\pi$) values~\cite{Ha06}: $a^{{}^3S_1}=5.4112(15)$~fm and
$a^{{}^1S_0}=-23.7148(33)$~fm.

The existence of these large scattering lengths leads to ``universal''
features in the dynamics of few-nucleon systems. These are a
consequence solely of the scale hierarchy $|a| m_\pi \gg 1$, and are
independent of details of the nuclear force. There has been much
recent progress on calculations in the universal effective theory
without explicit pions that describes this physics. This provides the
opportunity to compare precise, low-energy ($E < \frac{m_\pi^2}{M}$)
scattering experiments in the three- and four-nucleon systems with a
theory where calculations can be carried out with a comparable degree
of control. Progress in this direction in both theory and experiment
is discussed in Sec.~\ref{sec-pionless}.

At higher energies three-nucleon scattering experiments must be
analyzed using theories in which the pion is an explicit degree of
freedom. Recent experiments have
produced a wealth of data, including single- and double-polarization
observables, over a range of energies.  Ideally one would analyze these
data using nuclear forces derived from $\chi$PT.  The great benefit of
such an approach is the ability to derive three-nucleon (and, if
desired, four- and five- and six-...) nucleon forces within the same
framework, and to the same order in the chiral expansion, as is used
for the $NN$ force. This is discussed in Sec.~\ref{sec-3Nscattering}.

As with $n$-body forces, the operators that describe the interaction
of external probes (electrons, photons, pions, \ldots) with the
nucleus can also be expanded up to a chiral order that is consistent
with the order used to analyze the $NN$ potential. This ``chiral
effective theory'' ($\chi$ET) approach has now been applied to many
reactions in the $NN$ system, and calculations of reactions on the
tri-nucleons are just beginning.  These methods are particularly
valuable in facilitating the extraction of neutron properties from
experiments on multi-nucleon systems.  Neutron polarizabilities are
observables that have recently received particular attention in this
regard. Progress in these areas is summarized in
Sec.~\ref{sec-probes}. In addition to the topics discussed explicitly
there, related talks by F.~Myhrer on $pp \rightarrow pp \pi^0$ and
S.~Nakamura on the renormalization-group behavior of weak current
operators were also presented to the Working Group.

Finally, chiral perturbation theory is being used to analyze symmetry
breaking in the $NN$ system. New work on how parity-violating
operators appear in the $NN$ system provides a framework for the
analysis of a new generation of experiments that probe hadronic parity
violation. Isospin violation can also be included in the $NN$
potential using $\chi$PT techniques, and this allows a careful
examination of how isospin-violation effects appear in various
reactions. These issues are discussed in Sec.~\ref{sec-violation}.

\section{The $NN$ system: still fighting after all these years}

\label{sec-NN}

\subsection{Connection to lattice QCD}

One of the major developments in the $NN$ system since CD2003 is the
first $NN$ lattice calculation in full QCD~\cite{Beane:2006mx}. 
The NPLQCD collaboration
computed QCD correlation functions by using Monte Carlo techniques
to evaluate the QCD path integral on a discrete Euclidean
space-time lattice.  This starts directly from QCD, thereby allowing
truly ``{\em ab initio}'' calculations of nuclear physics
observables. However, the computational effort required for these
calculations increases sharply (i) with the number of valence quarks
involved and (ii) as the light quark masses are lowered to approach
physical values.  

S.~Beane reported on these results of the NPLQCD collaboration.
Their hybrid calculations used configurations
generated by the MILC collaboration including staggered sea quarks
and they evaluated the $NN$ correlator in this background using
domain-wall valence quark propagators. 
Applying the L\"uscher method, the spin-singlet and spin-triplet
S-wave scattering lengths were extracted at three
different pion masses between 350 and 590 MeV. At the lowest pion mass
the values $a^{{}^3S_1} = (0.63 \pm 0.74 \pm 0.2)$ fm and $a^{{}^1S_0} = (0.63 \pm
0.50 \pm 0.2)$ fm were found. At this pion mass the errors from a
chiral extrapolation to the physical $m_\pi$ are still large, but chiral
extrapolation from $m_\pi=139$ MeV shows that the result is
consistent with the experimental values, within (sizeable) error
bars.

Thanks to continuing advances in computer power
much progress in this direction can be expected over the next three
years. Lattice computations of $NNN$ correlators are a high priority
since they will provide access to three-nucleon forces directly from
QCD, but algorithmic and theoretical advances will be required in
order for such calculations to become a reality.

\subsection{Panel on power counting for short-distance operators}

Fifteen years after Weinberg's seminal papers on a chiral effective
theory description of nuclear physics, questions about this
technique remain.
The power counting (\ref{eq:V})
is quite successful phenomenologically, but various recent papers have
questioned its
validity~\cite{ES02,Nogga:2005hy,Birse:2005um}.
As part of the working group, we had a panel discussion with U. van
Kolck, U.-G. Mei\ss ner, and M. Pavon in order to try and shed some
light on this important issue.

The key problem is how to renormalize the Schr\"odinger equation with
(attractive) singular potentials. When the equation is solved there
are two acceptable wave-function solutions at short distances. The
``correct" linear combination must be fixed by physics input. This can
be implemented by adding a contact potential in such channels.  How
many contact terms are required when the $NN$ potential is computed to
a given order in the chiral expansion? So far this question has only
been intensively studied for the leading-order (LO) potential. That
the inclusion of one such contact term renormalizes the
${}^3$S$_1$-${}^3$D$_1$ channel at LO was established in
Refs.~\refcite{Fr99,Be01,PVRA05}. This vindicated Weinberg's counting
in that channel. However, Ref.~\refcite{Be01} confirmed that an
additional $m_\pi$-dependent contact term must be considered in the
${}^1$S$_0$ channel for renormalization there~\cite{Ka96}.  The issue
of how many contact terms to include in $V$ in other channels, where
Weinberg's counting does not indicate the presence of such a
contribution at LO, is still a controversial one.  Two main points of
view on this issue were presented in the panel.

U.~van Kolck argued that the necessity for the inclusion of contact
terms beyond those present in Weinberg's counting is dictated by strict
renormalization-group invariance of observables. Renormalization is
achieved when the effects of the cutoff $\Lambda$ are of the size of
higher-order terms in the $\chi$ET.  But when $\Lambda$ is varied over
a wide range, unphysical bound states appear in the $NN$ spectrum.
The only way to then guarantee renormalization-group invariance of $NN$
observables is to introduce additional contact terms which
renormalize $V$. Typically the necessity for these extra operators to
be included arises when $\Lambda$ is larger than the breakdown
scale of the theory, which is usually taken to be of order
$m_\rho$. For example, in Ref.~\refcite{Nogga:2005hy}
renormalization of the ${}^3$P$_0$ phase shift for energies up to 100
MeV is achieved for $\Lambda \geq 8$ fm$^{-1}$, but only after a contact
term that operates in this wave is included in the LO calculation.
M. Pavon's results support this point of view, although the
formulation employed in his work involved imposing boundary conditions
on the radial wave function at a very short distance.
He also used the requirement that wave functions
corresponding to different $NN$ energies should be orthogonal to
derive restrictions on the maximum number of contact terms that can be
included in a given channel.

The contrasting view was advocated by U.-G.~Mei\ss ner: that
calculations should only be carried out with a finite cutoff $\Lambda$
which is of the order of the expected breakdown scale of the $\chi$ET,
$m_\rho$.  Strict renormalization-group invariance is not required,
but the cutoff should be varied around $m_\rho$ in order to get a
lower bound on the size of omitted short-distance physics. There is no
point in considering $\Lambda \gg m_\rho$, since the error in the
calculation is not expected to decrease in this
regime\cite{Lepage:1997cs}.  In this case, the coefficients of
short-distance operators scale as predicted by naive dimensional
analysis with respect to $\Lambda$, and Weinberg's counting holds.  In
Ref.~\refcite{Epelbaum:2006pt} Epelbaum and Mei\ss ner examined the
behavior of $NN$ phase shifts and compared to the results of Nogga 
{\it et al}~\cite{Nogga:2005hy}. They found that a
result for the ${}^3$P$_J$ waves that deviates from the asymptotic
result of Ref.~\refcite{Nogga:2005hy} by less than the theoretical
uncertainty of a LO calculation could be achieved for $\Lambda \sim
600$ MeV.

During the discussion no consensus was achieved between the
panelists. Fundamental disagreement remained about whether it was
necessary to consider cutoffs $\Lambda\gg m_\rho$. Mei\ss ner argued
that Ref.~\refcite{Epelbaum:2006pt} showed that for all practical
purposes the range $\Lambda \lsim m_\rho$ contained all useful
information.  A. Nogga pointed out that it would not have been
possible to reach this conclusion---independent of data---in
Ref.~\refcite{Epelbaum:2006pt} if the cutoff-independent phase shifts
of the analysis of Ref.~\refcite{Nogga:2005hy} had not already
existed. Following this argument, for each new order in $V$, an
analysis which demands strict renormalization-group invariance may be
necessary before we can identify regions in $\Lambda$-space where
Weinberg's counting can be used in practice.

Mei\ss ner in particular, felt that no conclusions could be drawn
about power counting for short-distance operators until their role in
renormalization had been examined with $V$ calculated to orders beyond
leading.  Some of the contact terms promoted to LO in the study of
Ref.~\refcite{Nogga:2005hy} appear in the NLO potential, and (almost)
all of them are present by N$^3$LO. Therefore in N$^3$LO calculations
employing Weinberg's counting may be consistent with the requirement
of $\Lambda$-independence over a broad range. Such a resolution is
possible because, for high enough $J$, phase shifts can be calculated
in perturbation theory throughout the entire energy range for which
$\chi$PT is valid. However, it should also be noted that as higher
orders are considered the long-distance part of $V$ becomes even more
singularly attractive in some of the P- and D-waves discussed by the
panel. Whether this leads to further difficulties for Weinberg's power
counting remains to be seen. Indeed, issues associated with $V$ at NLO
and beyond can only be resolved by calculation, and the majority of
Working Group participants felt that studies which examine such issues are
very important.

\subsection{$\pi$NN coupling constant controversy resolved}

In the $NN$ system, many measurements have been performed in the last 30
years. In 1993 the Nijmegen group performed a partial-wave analysis
(PWA) of these data, and thereby obtained a statistically consistent
database of over 4000 $np$ and $pp$ data~\cite{St93}. In doing this
they found a (charged) $\pi NN$ pseudoscalar coupling of $\frac{g^2}{4
  \pi}=13.54 \pm 0.05$. This disagreed with a dedicated measurement of
backward-angle $np$ scattering at Uppsala, which yielded a result
consistent with the ``old'' higher value
$\frac{g^2}{4 \pi} \approx 14.4$~\cite{Uppsala}.

S.~Vigdor gave an overview of a recent experiment aimed at resolving
this discrepancy~\cite{Sarsour}. In this double-scattering experiment, the flux of
the neutron beam was accurately determined by counting the outgoing
protons in the reaction $^2$H($p,n$)2$p$ that produced the beam.  The
results of this absolute $np$ differential cross section measurement
at backward angles and a lab energy of 194 MeV are consistent with the
predictions of the Nijmegen PWA. The $NN$ database therefore now
provides an accurate and unambiguous determination of the ${\cal
  L}_{\pi N}^{(3)}$ LEC $b_{19}$, which is associated with the
Goldberger-Treiman discrepancy~\cite{GSS}.

\section{$p \ll m_\pi$: low-energy dynamics which isn't chiral}

\label{sec-pionless}

\subsection{N$^2$LO and beyond in the three-nucleon system}

One of the central goals of nuclear physics is to establish a
connection to QCD and its spontaneously broken chiral symmetry.  There
are some properties of nuclear systems that are due to chiral
dynamics.  Other properties, however, are simply a consequence of the
(apparently accidentally) large scattering lengths. Such properties
are ``universal'': they occur in a wide variety of systems from atomic
to nuclear to particle physics~\cite{BH05}, and can be studied in an
effective field theory (EFT) with contact interactions only. In nuclear
physics, this is a ``pionless'' theory, which can be applied to
processes with typical momenta $p\ll m_\pi$ and no external pions. It
allows for precise calculations of very low-energy processes and
exotic systems with weak binding such as halo nuclei.

L. Platter gave an overview of the pionless theory and showed various
applications from atomic and nuclear physics.  He presented results
from his recent calculations of $nd$ scattering and the triton that
included all effects up to second order in $p/m_\pi$ (i.e. N$^2$LO in
this EFT) using a subtractive renormalization scheme
\cite{Platter:2006ev,Platter:2006ad}. (For an overview of previous
higher-order calculations in the pionless theory see
Ref.~\refcite{BH05}.)  Platter's results imply that only the two-body S-wave
scattering lengths and effective ranges plus one datum from the
three-body system (usually taken to be $a_{nd}^{(1/2)}$ or the triton binding
energy) are needed to predict low-energy three-nucleon observables
with an accuracy $\sim 3$\%.  As a consequence, universal correlations
such as the Phillips and Tjon lines, as well as an analogous
correlation involving the triton binding energy and charge
radius~\cite{PH06}, persist up to N$^2$LO.

For precise higher-order calculations, it is important to know
at which order new three-body force terms appear 
\cite{Griesshammer:2005sj}.
A general classification of three-body forces in this theory
was presented by H. Grie\ss hammer. He also
illustrated the practical usefulness of the pionless theory by
showing precise predictions for a recent deuteron electrodisintegration
experiment at S-DALINAC~\cite{vNC02}, and threshold neutron-deuteron
radiative capture~\cite{Sadeghi2006}. 

\subsection{Experimental input and output}

Low-energy neutron-deuteron $(nd)$ scattering is characterized by two
scattering lengths: one in the quartet ($J=3/2$) and one in the
doublet ($J=1/2$) channel.  A novel method for measuring the
incoherent combination of these using polarized scattering was
outlined by O.~Zimmer.  His experiment obtains the incoherent $nd$
scattering length from the pseudomagnetic precession of neutrons in a
mixture of polarized deuterons and protons. Combining with the
well-known coherent scattering length yields an anticipated accuracy 
for $a_{nd}^{(1/2)}$ that is a factor of 10 better than
the present value $a_{nd}^{(1/2)}=0.65 \pm 0.04$
fm~\cite{vandenBrandt2004}. Since $a_{nd}^{(1/2)}$ is an input to
pionless EFT calculations in the three- and four-nucleon sectors this
experiment should also lead to more accurate predictions from that
theory.

With the ingredients of the calculations for the $NNN$ system under
control at low energies, theoretical attention is turning towards $4N$
systems.  The four-body system provides a fertile laboratory for
few-body physics since many features such as resonances and multiple
thresholds are present there, but are absent for $A<4$. $4N$ systems
are also the simplest ones where amplitudes of isospin T=3/2 can be
studied in the laboratory.  T. Clegg presented a talk on recent
measurements and plans for new measurements using the recently
developed polarized $^3$He target at TUNL in combination with
polarized proton and neutron beams.  In addition to analyzing powers,
spin-correlation coefficients have been measured. 

\subsection{The future}

The future of the pionless theory lies in the extension to larger
systems and in more precise higher-order calculations.  The new precise
$3N$ and $4N$ data from TUNL provide a challenge for the pionless
theory program. The best known example is the $A_y$ puzzle which is a
long-standing problem in few-nucleon physics.  There are dramatic
effects in the vector analyzing powers at low energies where the
pionless theory is applicable. As emphasized by T. Clegg,
these effects become more severe for
higher target masses which underscores the necessity for a better
understanding of the four-body system. Currently, only a leading-order
calculation is available and the general power counting of four-body
forces is not understood \cite{Platter:2004zs}.  As a consequence,
more theoretical work is required.  Many more interesting calculations
and experimental results pertinent to $4N$ systems can be expected
before the next Chiral Dynamics workshop.

The extension to systems with $A>4$ appears to be an opportunity for
lattice implementations of the pionless theory.  As discussed by
D.~Lee, the renormalization-group behavior of the three-body force was
already verified and lattice techniques have successfully been applied
to dilute neutron matter \cite{Borasoy:2005yc,Lee:2006vp}. 
These techniques can
also be extended to the theory with pions, and may provide novel ways
to compute nuclear bound states from $\chi$PT.

\section{Going higher in the $NNN$ system}

\label{sec-3Nscattering}

In the past fifteen years, experimental and theoretical developments
in the $NNN$ sector have taken place hand-in-hand. The numerical
solution of Faddeev equations describing three-body systems has
become routine, while high-precision experimental data have become
available thanks to a new generation of experimental facilities. These
include Bonn and Cologne in Germany, Kyushu in Japan, and TUNL in the
US for low energies, and KVI in the Netherlands, RIKEN and RCNP in
Japan, and IUCF in the US for intermediate energies.  In this section,
a summary of the measurements performed for the intermediate energies
will be presented.

J.~Messchendorp gave an overview of the experiments performed at KVI
in the past 10 years on the elastic and break-up channels in
proton-deuteron scattering. K.~Sekiguchi presented data from Japan on
the elastic and break-up channels in nucleon-deuteron
scattering~\cite{Ermisch}.  There is a persistent problem of
disagreement between the elastic $pd$ cross sections measured at KVI
and RIKEN at 135 MeV, but the bulk of data (including analyzing
powers) taken at both laboratories show unambiguously the need for an
additional ingredient in the theory beyond standard $NN$ potentials.
In some cases the addition of phenomenological $NNN$ forces brings the
results into very good agreement with the data but elsewhere,
particularly at higher energies, these three-nucleon forces do not
resolve the discrepancies between the data and the calculations.
Unfortunately, the $\chi$ET calculations at N$^3$LO 
required to accurately predict $NNN$ system observables for energies
larger than 100 MeV/nucleon have not yet been carried out, and only
calculations with phenomenological potentials are available for this
energy range.

In order to understand the spin structure of three-body forces,
double-scattering experiments with polarized deuteron beams have been
performed at both KVI and RIKEN. Spin-transfer coefficients from the
deuteron to the proton have been obtained with rather high precision
for a large part of the phase space~\cite{Sekiguchi}. These
measurements have been performed at lower energies, allowing a
comparison with the N$^2$LO predictions from $\chi$ET computed by
Witala and collaborators~\cite{Witala06}. Within the relatively large
error bands of the calculations there is good agreement with the data,
but some discrepancies remain.  The need to go to higher orders in
$\chi$ET is obvious.

The progress being made in this direction was described in the talk of
E.~Epelbaum. As explained above, the N$^3$LO $NN$ force already
exists, and gives results in quite good agreement with $NN$ data. Work
on deriving the (parameter-free) N$^3$LO $NNN$ force is ongoing.
Results that have been obtained so far for certain classes of diagrams
were presented. Completing the calculation of the full set of N$^3$LO
$NNN$ force diagrams and comparing N$^3$LO predictions with
experimental data will be a key test of $\chi$ET's usefulness in
three-nucleon systems, and we anticipate results by CD2009. Epelbaum
also outlined his derivation of a parameter-free $NNNN$ force, which
enters the nuclear potential at N$^3$LO~\cite{Ep06}. This is the first
microscopic computation of a $NNNN$ force, and estimates of its impact
on $\alpha$-particle binding suggest a 200--400 keV
effect~\cite{Ro06}.

Returning to experimental results, both Sekiguchi and Messchendorp
presented $pd$ break-up data, which showed the rich phase space
offered by this channel. Not surprisingly, the data supports the
inclusion of $NNN$ forces.  For the kinematics where the relative
energy of the two outgoing protons in the break-up reaction is small
one expects sensitivity to Coulomb effects. These effects have now
been explicitly included in the calculations of the Hanover-Lisbon
group~\cite{Deltuva06} and have been examined and shown to exist by
the Cracow-KVI measurements~\cite{Kistryn06}. Proton-deuteron break-up
data can thus be analyzed in a theoretical framework which makes it
clear which effects are due to the Coulomb interaction. Therefore, we
are no longer restricted to examining only $nd$ breakup data. This provides
a much richer $NNN$ database within which we can look for the
effects of chiral dynamics.

Messchendorp noted that the coupled-channels calculations of the
Hanover-Lisbon group, which treat the $\Delta(1232)$ as a dynamical
degree of freedom, do reasonably well when compared to the
measured data in all of the channels discussed above. This might
point to the fact that calculations of these observables in $\chi$ET
should include an explicit $\Delta(1232)$---especially if they wish to
address data at higher energies. This could improve the
convergence of $\chi$ET calculations, which at present show a sizeable
shift between NLO and N$^2$LO (c.f. Ref.~\refcite{Or96}).  This shift
signals the impact of two-pion-exchange in both the $NN$ and $NNN$
forces. As such it is a place where $\chi$ET gives concrete
predictions for the impact of two-pion-exchange physics on $nd$ and
$pd$ elastic and breakup data. More calculations that highlight the
impact of chiral physics on these data are needed.  

However, further opportunities for measurements of this physics appear
dangerously limited. The database in the $NNN$ system remains much
poorer than that in the $NN$ system, but many laboratories studying
this physics have shut down in the past few years and more will cease
operations in the near future.

\section{Soft Photons and Light Nuclei}

\label{sec-probes}

Electromagnetic reactions on light nuclei can also be calculated
within $\chi$ET. After using the chiral potential of Eq.~(\ref{eq:V})
to generate a wave function $|\psi \rangle$ for the nucleus---and, in
breakup reactions, a consistent wave function for the final scattering
state $|\psi_f \rangle$---one derives the current operator, $J_\mu$
for that system, again up to a given chiral order.  The matrix element
${\cal M} \equiv \langle \psi_f|J_\mu|\psi\rangle$ can then be
computed.

\subsection{Electron-deuteron scattering}

Work on current operators for the $NN$ system has been going on for
more than 10 years now.  Park, Min, Rho, and later collaborators have
computed deuteron photo- and electrodisintegration as well as weak
reactions using a hybrid approach where $J_\mu$ is sandwiched between
wave functions obtained from the AV18 potential (see,
e.g.~Refs.~\refcite{Park1994,Park2002}).  More recently, computations
of elastic electron-deuteron scattering using both $V$ and $J_\mu$
computed up to NLO~\cite{WM01} and N$^2$LO~\cite{Ph03,Ph06} in the
chiral expansion have been carried out. At N$^2$LO these calculations
seem to agree---within the combined theoretical and experimental error
bars---with the preliminary BLAST data presented in the talk of
R.~Fatemi.

\subsection{Compton scattering in $A=2$ and $A=3$}

Compton scattering from nucleons probes chiral dynamics in novel ways
that explore the interplay between the long-range pion cloud and
short-range operators in chiral perturbation theory~\cite{McG06}.
The spin-independent polarizabilities for the proton are now
quite well established, but the equivalent quantities for the neutron are
not as well constrained by existing experimental data. 
Better knowledge of $\alpha_n$
and $\beta_n$ would provide more information on how chiral
dynamics in Compton scattering expresses itself differently among the
two members of the nucleon iso-doublet.

H.~Grie\ss hammer reported on recent advances in calculations of
elastic $\gamma$d scattering. He and his collaborators have developed
a new power counting for $\gamma$d, designed for photon energies
$\omega \sim \frac{m_\pi^2}{M}$. In this regime,
the usual $\chi$PT counting cannot be applied to the $\gamma NN
\rightarrow \gamma NN$ operator, and resummation of certain classes of
diagrams is mandatory. When this resummation is complete the
calculation reproduces the Thomson limit for the zero-energy $\gamma$d
cross section~\cite{Hildebrandt2005}. This also reduces the theoretical
uncertainty in the calculation due to different assumptions about
short-distance physics.  Thanks to these
advances, as well as the inclusion of $\Delta(1232)$ effects that are
important at $\omega \sim 100$ MeV and backward angles,
$\alpha_n$ can now be extracted from the extant $\gamma$d
data with a precision of about 15\%~\cite{Hildebrandt2005}.

This is particularly exciting in light of the new data anticipated
from MAX-Lab. As described by K.~Fissum in his talk, an experiment
that is scheduled to begin there next summer will approximately quadruple
the world data on $\gamma$d scattering. Taken in concert with
theoretical advances this should facilitate an extraction of
spin-independent neutron polarizabilities at a precision comparable to
that at which $\alpha_p$ and $\beta_p$ are known.

D.~Choudhury reported on the first calculation of Compton scattering
from the Helium-3 nucleus. Her $\chi$ET calculation shows that there
is significant sensitivity to neutron spin polarizabilities in certain
$\gamma {}^3$He double-polarization observables. There are plans to
measure these at the HI$\gamma$S facility. Within Choudhury's NLO
calculation, a polarized ${}^3$He nucleus behaves, to a very good
approximation, as a polarized neutron. Neutron spin polarizabilities
can also be probed in $\vec{\gamma} \vec{\rm d}$ measurements at
HI$\gamma$S~\cite{CP05}, which provides an important cross-check on
our ability to calculate ``nuclear'' effects in Compton scattering.

Further progress in $\chi$ET calculations will reduce the theoretical
uncertainties that arise when neutron polarizabilities are extracted
from $\gamma$d or $\gamma {}^3$He data.  Meanwhile, accurate polarized
and unpolarized data on these reactions are anticipated from MAX-Lab
and HI$\gamma$S. With lattice QCD beginning to make
predictions~\cite{Lee1,Lee2} (albeit quenched ones) for baryon
polarizabilities there is an exciting opportunity for $\chi$ET to
connect experimental data from few-nucleon systems to the results of
lattice simulations.

\subsection{Photodisintegration}

A number of talks discussed photodisintegration of light nuclei as a
probe of nuclear forces. W.~Leidemann argued that this a
particularly attractive possibility, because the Lorentz Integral
Transform (LIT) is a technique by which bound-state methods can
be used to take a given $NN$ and $NNN$ force and obtain predictions
for the photodisintegration cross section.

The resultant predictions (with phenomenological force
models)~\cite{Gazit2006} for the photodisintegration of ${}^4$He
are in agreement with the data from Lund presented in the talk
of K.~Fissum~\cite{Lunddata}. They do not, however, agree with a
recent experiment at RCNP in Japan which used a novel
technique~\cite{Nagai}. P.~Debevec described an experiment that could
be done at HI$\gamma$S which, with careful control of systematics,
could pin down this cross section with very small error bars. Fissum
also anticipates a future experiment at MAX-Lab could significantly
reduce the error bars obtained in Ref.~\refcite{Lunddata}.

All of these experiments bear on the height of the first peak in the
$\gamma {}^4$He cross section. This quantity contains important
information on nuclear dynamics, and as such can be used to constrain
$\chi$ET (and perhaps even pionless) descriptions of nuclear forces. A
computation of ${}^4$He photodisintegration within $\chi$ET is
therefore a high priority, as is a definitive experiment to resolve
the discrepancies in the present ${}^4$He photoabsorption data.
Meanwhile, more detailed information on nuclear forces in general, and
the $\chi$ET $NNN$ force in particular, is expected from measurements
of three-body breakup (with neutron detection) of polarized ${}^3$He
by linearly-polarized photons. These experiments will take place at
HI$\gamma$S, and were described by X.~Zong.

\subsection{The Future}

Few-body methods such as the LIT, combined with $\chi$ET expansions for
the nuclear potential and electromagnetic current operators, provide
opportunities to confront $\chi$ET predictions with existing data
on electromagnetic reactions on light nuclei. But only a few
groups world-wide are performing such calculations, so this
opportunity may not be fully exploited. And as photon and electron
machines shut down or increase their energy, the possibility to use real
and virtual photons to probe chiral dynamics in light nuclei may
disappear. Over  the next few years the role of experimental facilities
such as MAX-Lab and HI$\gamma$S will be vital.

\section{Frontiers in symmetry breaking}

\label{sec-violation}

\subsection{Parity violation}

A discussion of hadronic parity violation experiments was presented by
S. Page. Various probes in few-body systems were
discussed. An experiment which is taking data presently is polarized
neutron radiative capture on protons at LANSCE. By measuring the very
small ($< 10^{-7}$) up-down $\gamma$-ray asymmetry, one can
constrain the low-energy constants that represent parity-violation in
the $\chi$ET~\cite{Zhu}, e.g. the parity-violating $\pi NN$ coupling
constant sometimes called $h_{\pi NN}$.

\subsection{Isospin violation}

The light-quark
mass difference $m_u-m_d$ is only a small fraction of the total mass
of the nucleon. But the large $NN$ scattering lengths magnify
this isospin-breaking to the point where it is a $\sim 10$\% effect in 
$a_{nn} - a_{pp}$.

Theoretical extractions from $pp$ data yield a strong proton-proton
scattering length of $a_{pp}=-17.3 \pm 0.4~{\rm
  fm}$~\cite{Wi95}. However, up until now, the neutron-neutron
scattering length, $a_{nn}$, has only been extracted from few-body
data. Different experiments on $nd$ breakup lead to numbers that
disagree, as described in the talk of C.~Howell. Over the next three
years we can anticipate new data on $a_{nn}$ from at least two
sources: a (hopefully) definitive $nd$ breakup experiment underway at
TUNL, and a re-analysis of LAMPF $\pi^- d \rightarrow nn \gamma$ data,
discussed in the talk of A. G\aa rdestig. In addition, the contribution of
V. Lensky pointed out that $\gamma d \rightarrow nn \pi^+$ could
provide a complementary $a_{nn}$ extraction. Once HI$\gamma$S attains
energies above pion threshold such an experiment could perhaps be done
there.

The NPLQCD collaboration has already examined the impact of $m_u -
m_d$ on the neutron-proton mass difference~\cite{Be06}. If this
calculation could be extended to $NN$ correlators, and the impact of
$m_u - m_d$ on $a_{nn} - a_{pp}$ predicted, it would allow us to
confront lattice results with high-precision extractions of the
scattering-length difference from few-nucleon systems.

\section{Conclusion}

All of this suggests an exciting future for Chiral Dynamics in these
systems. An era of calculations of few-nucleon bound states and
reactions that truly start from QCD is just beginning.  The
presentations at CD2006 allow us to foresee a future where lattice
simulations provide constraints on the low-energy constants that
appear in the $\chi$ET, and few-body methods allow us to completely
solve---even for systems with $A=4$ and beyond---that effective theory up
to a fixed order in the chiral expansion. The resultant computations,
which involve a mix of lattice, effective theory, and traditional
few-body techniques, can then be compared (including their theoretical
uncertainties!) with the wealth of experimental data in the $A=2,3$
and 4 sectors. We look forward to significant progress in the
formation of this linkage between the chiral dynamics of QCD and the
behavior of few-nucleon systems by the time of CD2009.

\section*{Acknowledgments}

We thank the conference organizers and group participants for their
efforts over the three afternoons of the Working Group. We also
acknowledge financial support from the US Department of Energy
(DE-FG02-93ER40756, DP), Deutsche Forschungsgemeinschaft (SFB/TR-16,
HWH), and Bundesministerium f\"ur Bildung und Forschung (Grant
no. 06BN411, HWH).

\end{document}